# Why the Atlantic meridional overturning circulation may not collapse

by Hans van Haren

Royal Netherlands Institute for Sea Research (NIOZ), P.O. Box 59, 1790 AB Den Burg, the Netherlands.
e-mail: hans.van.haren@nioz.nl

The extent of anthropogenic influence on the Earth's climate warrants studies of the ocean as a major player. The ocean circulation is important for transporting properties like heat, carbon and nutrients. A supposed major conduit is the Atlantic Meridional Overturning Circulation (AMOC). Schematically, it transports heat from the equator to the poles near the surface and carbon in the abyssal return[1]. As the AMOC is a complex nonlinear dynamical system, it is challenging to predict its potential to collapse from a statistical viewpoint[2] using a particular estimator of sea-surface temperature. Whilst this might be robust mathematically, it lacks physical insight of the drivers of the AMOC. As is argued below, physical processes will alter the estimators, and thereby statistical analyses.

In contrast with the atmosphere, the ocean is not a heat engine[3]. As a result, the AMOC is not buoyancy-driven via deep dense-water formation near the poles, which notably occur in sporadic pulses rather than continuously. Instead, the AMOC is wind- and tide-driven[3], with turbulent mixing by internal wave breaking being the important physics process. Winds, near the ocean surface, and tides, via interaction with seafloor topography deeper down, contribute about equally to generate internal waves that are found everywhere in the ocean interior. Such waves break predominantly at ubiquitous underwater seamounts and continental rises.

Without turbulent mixing, AMOC would be confined to a 100-m thick near-surface layer and the deep-ocean would be a stagnant pool of cold water[3]. However, the solar heat is mixed from the surface downward and the ocean is stably stratified in density all the way into its deepest trenches, as has been shown in hydrographic deep-ocean observations. Although turbulent mixing by internal wave breaking is insufficient in the ocean interior to maintain the stratification, such breaking is more than sufficient along ocean boundaries[4,5]. Especially large internal wave breaking occurs above steeply sloping topography[6-8]. As there are more and larger seamounts than mountains on land, equally abundant sloping seafloors lead to abundant turbulent mixing.

Recent detailed observations and numerical modeling have revealed the extent of internal tide breaking processes above ocean topography[9-11]. Quantification of the turbulent mixing shows that it occurs in bursts when highly nonlinear internal waves propagate as turbulent bores up a slope, once or twice a tidal cycle. Between bores, the turbulent mixing varies by an order of magnitude in intensity, with effects extending about 100 m vertically and several kilometers horizontally from the seafloor. Although intermittently occurring at a given position of the sloping seafloor and about 10% varying in arrival time, the turbulence is forced by the tide and winds in a stratified environment. The turbulent bores also resuspend sediment and thereby replenish nutrients away from the seafloor.

Question is whether the intensity of internal-wave induced deep-ocean turbulence is affected by variations in sea-surface temperature, with what consequences for AMOC.

Any variation to the nonlinear system of ocean circulation may encounter several complex feedback mechanisms, of which the effects are not yet fully understood. Although stable stratification hampers vertical turbulent exchange, it does not block the turbulent mixing. While stratification supports internal waves, turbulent mixing by internal wave breaking may decrease or destroy the stratification locally in time and space. However, a subsequent internal wave-phase will restratify the mixed patch, thereby maintaining its own support of stable stratification. Such a feed-back system may be at work when the ocean absorbs more heat.

Increased sea-surface temperature may lead to increased stratification, which may lead to less vertical turbulent exchange and more internal waves through the extension of their spectral band. As particular internal waves can propagate deep into the ocean interior, they can cause enhanced turbulent mixing elsewhere. Limited observations have thus far not provided evidence for an inverse correspondence between variations in turbulent mixing and variations in temperature across the near-surface photic zone along a longitudinal section of the NE Atalantic[12]. This lack of correspondence suggests a feedback mechanism at work mediating potential physical environment changes so that global warming may not affect vertical turbulent fluxes of heat, and thereby also of, e.g., carbon. More evidence is urgently needed, also from abyssal ocean observations.

While the anthropogenic influence on the Earth's climate is without doubt, the impact on the ocean circulation is not fully known because we lack sufficient information of the relevant processes that we cannot model yet. Therefore, we should be cautious in making predictions on future ocean circulation based on estimators that are uncertain proxies. Because no observational[12] or modeling[13] physics evidence exists that sea-surface temperature is a solid estimator of AMOC-strength variations, other properties like salinity, besides temperature contributing to density variations in the ocean, density gradients (stratification), and turbulence intensity may be considered.

As for the ocean circulation in the horizontal plane near its surface with most impact on mankind, wind will remain the main driver. As long as the Earth rotation does not alter direction, wind will maintain its general course. The atmosphere remains a key player in the global heat transport. Simultaneously, the importance of stratification and turbulent mixing cannot be underestimated for life near the ocean-surface and in the deep, as without these processes it will come to a halt.

**Competing interests** The author declares no competing interest.

**Acknowledgments** I thank L. Gerringa for commenting a previous draft of the manuscript.